\documentclass[pra,twocolumn,amsmath,amssymb,letterpaper,superscriptaddress,longbibliography]{revtex4-2} 
\usepackage{graphicx}
\usepackage{amsmath}
\usepackage{amsfonts}
\usepackage{amssymb}
\usepackage{braket}
\usepackage{natbib}
\usepackage[colorlinks,linkcolor=blue,filecolor=blue,urlcolor=blue,citecolor=blue]{hyperref}

\begin{document}

\title{Encircling the Liouvillian exceptional points: a brief review}

\author{Konghao Sun}
\affiliation{CAS Key Laboratory of Quantum Information, University of Science and Technology of China, Hefei 230026, China}
\author{Wei Yi}
\email{wyiz@ustc.edu.cn}
\affiliation{CAS Key Laboratory of Quantum Information, University of Science and Technology of China, Hefei 230026, China}
\affiliation{Anhui Province Key Laboratory of Quantum Network, University of Science and Technology of China, Hefei 230026, China}
\affiliation{CAS Center For Excellence in Quantum Information and Quantum Physics, Hefei 230026, China}

\begin{abstract}
Exceptional points are the branch-point singularities of non-Hermitian Hamiltonians, and have rich consequences in open-system dynamics.
While the exceptional points and their critical phenomena are widely studied in the non-Hermitian settings without quantum jumps, they also emerge in open quantum systems depicted by the Lindblad master equations, wherein they are identified as the degeneracies in the Liouvillian eigenspectrum.
These Liouvillian exceptional points often have distinct properties compared to their counterparts in non-Hermitian Hamiltonians,
leading to fundamental modifications of the steady states or the steady-state-approaching dynamics.
Since the Liouvillian exceptional points widely exist in quantum systems such as the atomic vapours, superconducting qubits, and ultracold ions and atoms, they have received increasing amount of attention of late.
Here we present a brief review on an important aspect of the dynamic consequence of Liouvillian exceptional points, namely the chiral state transfer induced by the parametric encircling the Liouvillian exceptional points.
Our review focuses on the theoretical description and experimental observation of the phenomena in atomic systems that are experimentally accessible.
We also discuss the on-going effort to unveil the collective dynamic phenomena close to the Liouvillian exceptional points, as a consequence of the many-body effects therein.
Formally, these phenomena are the quantum-many-body counterparts to those in classical open systems with nonlinearity, but hold intriguing new potentials for quantum applications.
\end{abstract}

\maketitle
\section{Introduction}
In a typical non-Hermitian system, the dynamics is effectively driven by a non-Hermitian Hamiltonian~\cite{nhrev}.
As a common feature of a wide class of non-Hermitian matrices, multiple eigenvectors and eigenvalues of a non-Hermitian Hamiltonian can simultaneously coalesce at certain critical points in the parameter space.
These critical points, known as the exceptional points (EPs), correspond to the so-called branch-point singularities in the eigenspectrum, where the Hamiltonian cannot be diagonalized~\cite{ep1966,ep1998,ep1999}.
As such, the EPs are fundamentally different from conventional degeneracies in Hermitian systems.
In recent years, the EPs have been widely discussed in the context of non-Hermitian models with the parity-time symmetry~\cite{pt1,pt2,pt5,pt6,pt7,ep3,ep4,eptopo,ep5}, where they appear as the demarcation between a spectral region with entirely real eigenvalues (dubbed the parity-time unbroken regime), and one without (dubbed the parity-time broken regime).
Being the transition point between these spectrally and dynamically distinct regions, the EPs have been extensively studied in the context of parity-time symmetric models in connection with phenomena such as power oscillations~\cite{pow1,pow2,pow3}, directional transport~\cite{dir1,dir2} and lasing~\cite{las1,las2,las3,las4}.
However, EPs also exist in systems without the parity-time symmetry, and are therefore more general~\cite{nhrev,ep3}.
Thanks to the singular nature of EPs, a system exhibits many interesting behaviors in their vicinity, including universal criticality~\cite{crit1,crit2,crit3,crit4}, non-reciprocal dynamics~\cite{ec1,ec2,ec3,ec4,ec5,cla3,cla1,cla2,phot1,ions,nv,ca}, enhanced entanglement generation~\cite{ent1,ent2}, and strong sensitivity to external perturbations~\cite{sen1,sen2,sen3}.
These properties have been confirmed in a wide range of classical and quantum mechanical systems, including optics and photonics~\cite{ec3,ec4}, optomechanics~\cite{opm1,opm2} acoustics~\cite{acou1,acou2,acou3}, atomic gases~\cite{luoleexp,ca}, and superconducting qubits~\cite{sq1,sq2, ent2}.
Understanding EPs and the related phenomena create opportunities for developing useful applications in terms of sensing and topological transport.

In the quantum regime, EPs, and non-Hermitian physics in general, are mostly discussed in the context of conditional dynamics of an open quantum system. Specifically, the full quantum dynamics of an open system is governed by the Lindblad master equation under the Markov approximation~\cite{lindblad}. In the quantum trajectory picture, the density matrix evolution can be unravelled as an ensemble of quantum trajectories~\cite{traj}. The quantum state in each trajectory undergoes evolution driven by an effective non-Hermitian Hamiltonian, but is interrupted by stochastic quantum jumps. It follows that, when considering the transient dynamics or by selecting trajectories without quantum jumps (hence conditioned), the dynamics is governed purely by the non-Hermitian Hamiltonian~\cite{nhrev}. While such a scheme can be experimentally implemented through post selection, the practice is limited to single-particle systems, or non-interacting ones where quantum statistics is not essential.

\begin{figure}[tbp]
	\centering
	\includegraphics[width=0.9\linewidth]{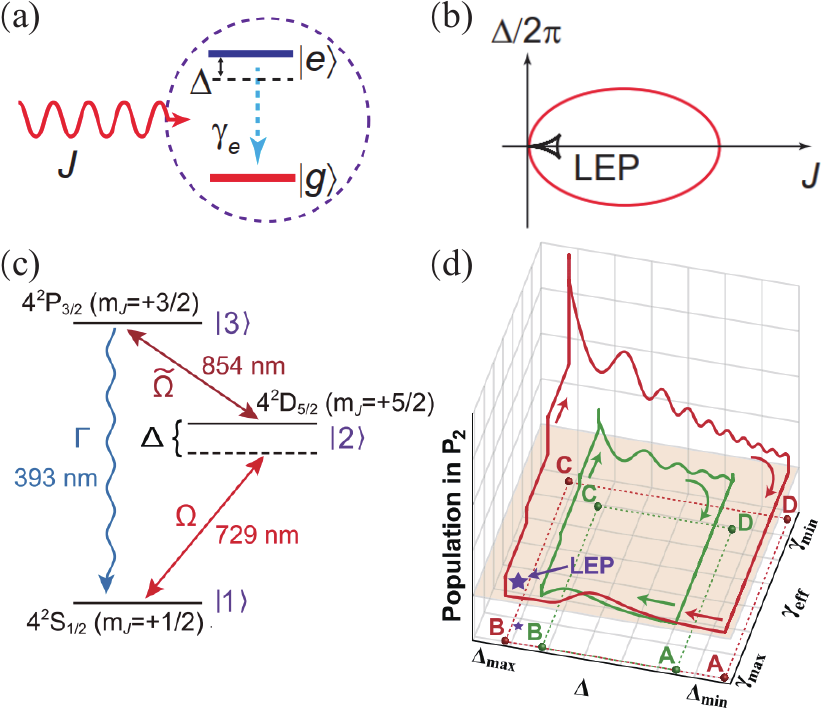}
	\caption{(a) Schematics of LEP engineering in the superconducting qubit experiment~\cite{sq2}. (b) The corresponding Liouvillian exceptional structure in the parameter space of Ref.~\cite{sq2}. (c) Level scheme for generating LEP in the trapped ion experiment~\cite{qhe2}. (b) The corresponding LEP in the parameter space of ~\cite{qhe2}. Here (a)(b) are adapted from Ref.~\cite{sq2}, copyright by the American Physical Society.
And (c)(d) are adapted from Ref.~\cite{qhe2}, copyright by the Americal Physical Society.}
	\label{ep_experiment}
\end{figure}

On the other hand, the Lindblad master equation, complete with full quantum jump processes, describes the evolution of the density matrix under the Liouvillian superoperator. Since the Liouvillian itself can be represented as a non-Hermitian matrix acting on a vectorized density matrix, the EPs can also arise in the Liouvillian spectrum~\cite{bloch_lep,LEP1,LEP2,hlep_encircling}, leading to unique dynamics in the full quantum dynamics of the master equation. Along this line of thinking, the Liouvillian EPs are defined and studied in recent theoretical and experimental works~\cite{hlep_encircling,LEP1,LEP2,hlep_encircling,qhe,qhe2,sq1,sq2,sunyi}.
In Fig.~\ref{ep_experiment}, we illustrate two recent experiments where LEPs were engineered in superconducting qubits and trapped ions, respectively.
It is found that, the Liouvillian EPs have distinct impact on the open-system dynamics, compared to their counterparts in non-Hermitian Hamiltonians.
More importantly, since the application of the quantum master equation is not limited to non-interacting systems or transient dynamics, the Liouvillian EPs can become relevant in open systems with many-body correlations~\cite{wangopen,zhaiopen,yiopen,cavityskin}.
Thus, understanding Liouvillian EPs in a generic open quantum system exemplifies the advancing frontier of non-Hermitian physics, where
the wealth of non-Hermitian phenomena is no longer restrained to the classical or semi-classical regime, but lends meaningful insights to quantum many-body processes.

In the following, we provide a brief review on the study of Liouvillian EPs, focusing on the experimentally relevant atomic systems.
We first review conventional EPs in non-Hermitian Hamiltonians, using a simple two-level system as an example.
Then, by adding different quantum jump operators, we discuss how the density-matrix dynamics admits a non-Hermitian description, as well as the emergence of Liouvillian EPs in the framework of Lindblad master equations.
As the main content of the review, we give some concrete examples, all experimentally related, to the role of Liouvillian EPs in open-system dynamics, and show how non-Hermitian physics emerge within the framework of master equations~\cite{ca,sq1,sq2,qhe,rydbergexp}.
Finally, we discuss a recent experiment on the collective topological state transfer near a Liouvillian EP, where both non-Hermiticity and many-body effects play a key role.


\section{Hamiltonian EP and Liouvillian EP}
To illustrate the connection and difference between the EPs in non-Hermitian Hamiltonians and those of the Liouvillians, we first consider an exemplary two-level model. The Hamiltonian in the corresonding spin basis can be written as
\begin{align}
H=J\sigma_x-i\frac{\Gamma}{2}\sigma_z,\label{eq:Hpt}
\end{align}
where $J$ and $\Gamma$ are positive coefficients. The Hamiltonian possesses the parity-time symmetry, with $(PT)H(PT)^{-1}=H$, where the parity operator $P=\sigma_x$, and the time-reversal operator $T$ corresponds to complex conjugation. The eigenvalues of the Hamiltonian are straightforward to calculate
\begin{align}
E_{\pm}=\pm\sqrt{J^2-\frac{\Gamma^2}{4}}.
\end{align}
Here the eigenvalues are real for $J\geq \Gamma/2$, and imaginary for $J<\Gamma/2$. Clearly, $J=\Gamma/2$ corresponds to the aforementioned EP, separating the parity-time unbroken and broken regimes. To understand the criticality of the EP, we examine the eigenvectors for $J\geq \Gamma/2$, with
\begin{align}
|\psi_{\pm}\rangle=\frac{\sqrt{2}}{2J}
\left(
\begin{array}{c}
-i\frac{\Gamma}{2}\pm \sqrt{J^2-\frac{1}{4}\Gamma^2}\\ J
\end{array}\right).
\end{align}
Notably, the eigenvectors coalesce at the EP $J=\Gamma/2$, where the non-Hermitian matrix cannot be diagonalized.
The ill-conditioned matrix leads to many intriguing observable phenomena, and have stimulated much research interest.

Likewise, exceptional structures also exist in the Liouvillian eigenspectrum. Based on Hamiltonian (\ref{eq:Hpt}), we consider the following Lindblad master equation (we set $\hbar$=1)
\begin{align}
\dot{\rho}=\mathcal{L}\rho=-i[H,\rho]+L\rho L^\dag-\frac{1}{2}L^\dag L\rho -\frac{1}{2}\rho L^\dag L,
\end{align}
where $H=J\sigma_x$, and the quantum-jump operator $L=\sqrt{\Gamma} \sigma_-$. Note that when we drop the recycling term $L\rho L^\dag$, the remaining equation of motion describes the evolution of the density matrix by a non-Hermitian effective Hamiltonian $H_{\text{eff}}=H-i\frac{\Gamma}{2}L^\dag L$, which reproduces Hamiltonian (\ref{eq:Hpt}).
As we discuss in the following section, neglecting the recycling term is referred to as the no-jump condition, also known as the semi-classical limit or the post-selection condition. Though with limitations, it offers a convenient and practical route, connecting the dynamics of quantum open systems with those under a non-Hermitian effective Hamiltonian.

More relevant to the discussion here, we vectorize the density matrix
\begin{align}
	\rho=\begin{bmatrix}
	    \begin{array}{cc}
	    	\rho_{11} & \rho_{12}\\
	    	\rho_{21} & \rho_{22}
	    \end{array}
	\end{bmatrix}
    \rightarrow
    \begin{pmatrix}
    	\rho_{11}\\
    	\rho_{12}\\
    	\rho_{21}\\
    	\rho_{22}
    \end{pmatrix},
\end{align}
and write the Liouvillian superoperator in a non-Hermitian matrix form
\begin{align}
	\mathcal{L}=
	\begin{bmatrix}
		\begin{array}{cccc}
			0 & iJ  & -iJ & \Gamma \\
			iJ &  -\frac{\Gamma}{2} & 0 & -iJ \\
			-iJ & 0 & -\frac{\Gamma}{2} & iJ\\
			0 & -iJ & iJ & -\Gamma\\
		\end{array}
	\end{bmatrix}.\label{eq:Lmatrix}
\end{align}
Importantly, now that the Liouvillian is represented by a non-Hermitian matrix, it can host exceptional structures. The exceptional points in the Liouvillian eigenspectrum are referred to as the Liouvillian exceptional points (LEPs), in contrast to the Hamiltonian exceptional points (HEPs). Specifically, the Liouvillian eigenspectrum $\lambda$ is calculated through $\mathcal{L}\rho_\lambda=\lambda\rho_\lambda$, with the corresponding Liouvillian eigenstates $\rho_{\lambda}$. For the Liouvillian in (\ref{eq:Lmatrix}), we then have
\begin{align}
\lambda_1=0,\,\lambda_2=-\frac{\Gamma}{2},\,\lambda_{3,4}=\frac{1}{4}\left(-3\Gamma\pm \sqrt{\Gamma^2-64J^2}\right).
\end{align}
While the eigenvalue $\lambda_1=0$ corresponds to the steady-state solution, $\lambda_{3,4}$ coalesce at $\Gamma=8J$ and $\lambda_{2,3}$ coalesce at $J=0$, representing second-order LEPs. These LEPs lie away from the steady-state solution, suggesting that their impact is in the intermediate time scales, before the system relaxes to the steady state at long times. Note that, equivalent to the treatment above, one may also write down the corresponding optical Bloch equations, and extract information of the LEPs from the coefficient matrix~\cite{bloch_lep}.

\begin{figure}[tbp]
	\centering
	\includegraphics[scale=1]{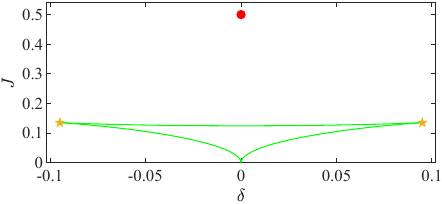}
	\caption{The Liouvillian exceptional structure of Eq.~(\ref{eq:Lmatrix2}) on the $\delta$--$J$ plane.
The green lines are the second-order Liouvillian exceptional lines, the yellow stars indicate the third-order LEP. The red point is the second-order HEP of the corresponding non-Hermitian Hamiltonian under the no-jump condition. We take $\Gamma=1$ as the unit of energy for our calculations.}
	 \label{review1}
\end{figure}

More generally, the connection and distinction between the LEP and HEP have been systematically discussed in Refs.~\cite{LEP1,LEP2}.
Although LEPs and HEPs have essentially different properties, an open system
can have exceptional structures that are the direct correspondence to a HEP of the non-Hermitian Hamiltonian under the no-jump condition.
This is the case with the Liouvillian above. The Liouvillian eigenspectrum features an exceptional point, but shifted in parameters compared to those of the HEP of the non-Hermitian Hamiltonian.

Furthermore, when a detuning term $\frac{\delta}{2}\sigma_z$ is added to the Hermitian Hamiltonian $H$, the matrix form of the Liouvillian operator becomes
	\begin{align}
		\mathcal{L}=
		\begin{bmatrix}
			\begin{array}{cccc}
				0 & iJ  & -iJ & \Gamma \\
				iJ &  -\frac{\Gamma}{2}-i\delta & 0 & -iJ \\
				-iJ & 0 & -\frac{\Gamma}{2}+i\delta & iJ\\
				0 & -iJ & iJ & -\Gamma\\
			\end{array}
		\end{bmatrix}.\label{eq:Lmatrix2}
	\end{align}
The corresponding Liouvillian eigenspectrum shows exceptional lines on the $\delta$--$J$ plane, ending at two third-order exceptional points as illustrated in Fig.~\ref{review1}.
As such, the HEP of the non-Hermitian Hamiltonian develops into an exceptional structure in the Liouvillian eigenspectrum, consisting of exceptional lines and higher-order exceptional points.

\section{Dynamic consequence of EPs}
One outstanding feature of the EPs is the sensitivity of eigenvalues to external perturbations. For instance, by adding a weak perturbative term $\epsilon \sigma_x$ to Hamiltonian (\ref{eq:Hpt}), the eigenvalues at the EP become
\begin{align}
E_{\pm}=\pm \sqrt{\epsilon(2J+\epsilon)},
\end{align}
splitting by a small amount $\sim \epsilon^{\frac{1}{2}}$. More generally, at a higher-order EP, where more than two eigenvectors and eigenvalues coalesce, the splitting is of the order $\epsilon^{\frac{1}{n}}$. Such a sensitivity is the basis for discussions of EP-enhanced sensing, which have been extensively reported and reviewed in recent years~\cite{ep3,ep4,sen1,sen2,sen3}.

\begin{figure*}[tbp]
	\centering
	\includegraphics[width=1\textwidth]{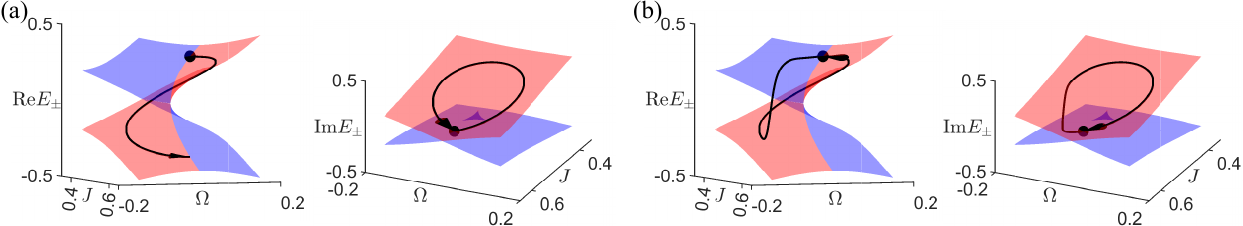}
	\caption{Trajectories along the Riemann surface corresponding to the non-Hermitian model (\ref{eq:encircle}). The Riemann surface is colored red (blue), indicating the gain (loss) of the eigenvalues. Parameters change in (a) counterclockwise and (b) clockwise directions end up with different states after one cycle. For numerical calculations, we take the parameters $\Gamma=1$, $r=0.1$, and $T=100$.}
	\label{review2}
\end{figure*}

Here we focus on a general dynamic consequence, which derives from
the complex spectral topology in the parameter space near an EP. Take, for instance, a parameterized non-Hermitian model based on Eq.~(\ref{eq:Hpt})
\begin{align}
H(t)=J(t)\sigma_x-[\Omega(t)+i\frac{\Gamma}{2}]\sigma_z,\label{eq:encircle}
\end{align}
where $J(t)=\Gamma/2+r\cos(\omega t)$, $\Omega(t)=r\sin(\omega t)$, and $t$ is understood as a parameter but would eventually parameterize the time dependence of the variables. The eigenvalues are
\begin{align}
E_{\pm}=\pm\sqrt{r^2+\Gamma r e^{i\omega t}},
\end{align}
which, in the parameter space of $(J,\Omega)$, lead to the Riemann sheets illustrated in Fig.~\ref{review2}. Here an EP exists at $(J=\Gamma/2, \Omega=0)$, which represents the endpoint of the branch cut along $\Omega=0$ with $J\leq 0$. As such, the EP is usually referred to as the branch-cut singularity of the non-Hermitian matrix.

Such a geometry has interesting dynamic consequences. Consider the evolution of an eigenstate under $H(t)$ where $t$ now represents time. Intuitively, when the rate of time variation is slow compared to the real energy gap along the path (roughly with $\frac{2\pi}{\omega}\text{min}(|E_+-E_-|)\gg 1$), the time-evolved state should adiabatically follow the instantaneous right eigenstate of $H(t)$
\begin{align}
|\psi_{\pm}\rangle\propto
\left(
\begin{array}{c}
-[\Omega(t)+i\frac{\Gamma}{2}]\pm \sqrt{J(t)^2+(\Omega(t)+i\frac{\Gamma}{2})^2}\\ J(t)
\end{array}\right).\label{eq:psipm}
\end{align}

We track such a process with the trajectory $\overline{E}(t)$ on the Riemann sheet, with
\begin{align}
\overline{E}(t)=\frac{\sum_{i=\pm}|\langle \chi_i(t)|\psi(t)\rangle|^2 E_i(t)}{\sum_{i=\pm}|\langle \chi_i(t)|\psi(t)\rangle|^2},
\end{align}
where $|\chi_i(t)\rangle$ is the left eigenstate of the instantaneous Hamiltonian, defined as $H^{\dagger}(t)|\chi_i(t)\rangle=E^\ast_i(t)|\chi_i(t)\rangle$.
Importantly, when the trajectory crosses the branch cut, the band indices ($\pm$) of the right eigenstate switch, due to the sign change in the front of the square root in Eq.~(\ref{eq:psipm}).
Correspondingly, assuming the system is initialized in an eigenstate, under one full cycle of the parameter change, it undergoes an eigenstate switching under the adiabatic condition. This is illustrated in Fig.~\ref{review2}(a), where the parameters change in a counterclockwise direction along the trajectory.

However, the adiabatic condition does not always hold. As illustrated in Fig.~\ref{review2}(b), when the parameters change in a clockwise direction, the system comes back to the original eigenstate, due to a non-adiabatic jump that switches the eigenstate one more time along the path. Interestingly, such a non-adiabatic jump occurs even if the condition $\frac{2\pi}{\omega}\text{min}(|E_+-E_-|)\gg 1$ is satisfied. Such a phenomenon derives from the non-Hermiticity of the system.
Specifically, expanding the time-evolved state onto the basis of the instantaneous right eigenstates~\cite{ec5}
\begin{align}
|\psi(t)\rangle=\sum_{n=\pm} c_n(t)|\psi_n(t)\rangle,
\end{align}
the coefficients satisfy
\begin{align}
i\hbar\frac{\partial}{\partial t}c_n=[E_n(t)-i\hbar \langle\chi_n|\frac{\partial}{\partial t}|\psi_n\rangle]c_n-i\hbar \langle\chi_n|\frac{\partial}{\partial t}|\psi_{-n}\rangle c_{-n}.
\end{align}
The last term on the right-hand side describes the non-adiabatic transition between different eigenstates. The term can have significant impact when an unoccupied eigenstate has an eigenvalue with a positive imaginary component (or simply a less negative imaginary component), leading to an exponentially increased likelihood of a non-adiabatic jump into the said state. As a result, depending on the initial state and the encircling direction, the system undergoes an eigenstate switch only along one direction of the parameter change. This is referred to as the chiral state transfer in the literature~\cite{ep3,ep4}. While the process of the parameter change is known as the EP encircling, the path of the parameter change does not have to encircle the EP: the chiral state transfer can occur when the path lies close to the exceptional structure.
Exceptions can arise under special circumstances, where the chiral state transfer can occur on trajectories far way from any EPs~\cite{ca,cla3}.

\section{Implementing non-Hermitician Hamiltonian in quantum systems}

The chiral state transfer under the Hamiltonian EP encircling has been experimentally demonstrated in a variety of physical systems~\cite{cla1,cla2,phot1,ions,nv,ca,sq1,sq2}. A recent experiment further establishes the phenomenon in an ultracold gas of fermions~\cite{ca}. While in all cases, the dynamics are driven by a non-Hermitian effective Hamiltonian, in quantum systems, the no-jump condition leading to the non-Hermiticity is typically imposed through post selection.

\begin{figure}[htbp]
	\centering
	\includegraphics[width=0.48\textwidth]{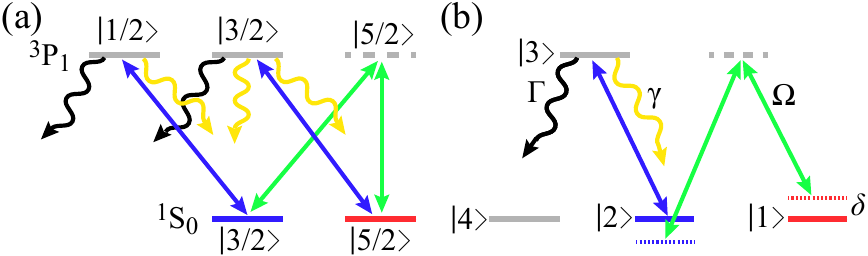}
	\caption{(a) The level scheme in Ref.~\cite{ca}, where states in the $^1S_0$ and $^3P_1$ manifolds of ultracold $^{173}$Yb atoms are used.
 (b) Level scheme for a simplified four-level model.
	} \label{review3}
\end{figure}

Take the cold-atom experiment for example. Ultracold $^{173}$Yb atoms are prepared in the $|m_F=3/2\rangle$ and $|m_F=5/2\rangle$ states of the ground-state $^1S_0$ ($F=5/2$) manifold. The two states are coupled through a two-photon Raman process, and the state $|m_F=3/2$ is further laser coupled to an electronically excited state in the $^3P_1$ ($F=7/2$) manifold.
This introduces a controlled atom loss to the system, due to the finite linewidth of the excited state. The level scheme is illustrated in Fig.~\ref{review3}(a).
Neglecting for now the spontaneous emission back to the $\{|m_F=3/2\rangle, |m_F=5/2\rangle\}$ subspace, one assumes that the atoms undergo the spontaneous emission process are lost from the system. Under such a condition, the dynamics of the remaining atoms are governed by a non-Hermitian effective Hamiltonian
\begin{align}
H_{\text{eff}}=\frac{\delta}{2}\sigma_z-\Omega\sigma_x-i\frac{\Gamma}{4}(1-\sigma_z),\label{eq:coldatomH}
\end{align}
where $\Omega$ and $\delta$ are respectively the Rabi frequency and detuning of the Raman process, and $\Gamma$ is the laser-induced loss rate. For the convenience of discussion, we label
$|1\rangle=|m_F=5/2\rangle$ and $|2\rangle=|m_F=3/2\rangle$, and define the Pauli operators such that $\sigma_z=|1\rangle\langle 1|-|2\rangle\langle 2|$. We also label the excited state used for the laser-induced loss as $|3\rangle$, as illustrated in Fig.~\ref{review3}(b).

From the perspective of quantum open systems, the non-Hermitian effective Hamiltonian above can be derived by imposing a no-jump condition on the quantum master equation
\begin{align}
	\dot{\rho}=-i[H,\rho]+L_{1}\rho L_{1}^{\dagger}-\frac{1}{2}L_{1}^\dag L_{1}\rho-\frac{1}{2}\rho L_{1}^\dag L_{1}, \label{eq1}
\end{align}
where $H=\frac{\delta}{2}\sigma_z-\Omega\sigma_x$, and $L_1=\sqrt{\Gamma}|4\rangle\langle 2|$ describes the laser-induced loss from state $|2\rangle$ to a bystander state $|4\rangle$. The no-jump condition corresponds to neglecting the recycling term $L_1\rho L_1^\dag$, which applies when one focuses only on the subspace $\{|1\rangle,|2\rangle\}$. Equivalently, from all possible time-evolved states, one selects out only those with no support on $|4\rangle$. In practice, this is natural if one assumes that atoms in $|4\rangle$ are no longer trapped while only those remain trapped are detected. Since the atoms that remain in the $\{|1\rangle, |2\rangle\}$ subspace necessarily have not gone through the quantum jump (spontaneous emission) to state $|4\rangle$,
the non-Hermitian description is therefore always valid within the subspace. The exponential decrease in the modulus square of the wave function directly corresponds to the exponential decay of the total atom number within the subspace.
It is worth pointing out that such a natural implementation of post selection (by detection) only applies when the atoms are non-interacting. Even as the recycling term mixes density matrices from sectors with different atom numbers, the overall density matrix in each sector is similar in structure, featuring a direct product of those of individual atoms. This is not the case in the presence of interactions, under which density matrices in different atom-number sectors are coupled through the interplay of dissipation and interaction.

Under the condition above, the master equation reduces to the non-Hermitian description (\ref{eq:coldatomH}), which, having the same structure as Hamiltonian (\ref{eq:encircle}), forms the basis of the observed EP encircling in Ref.~\cite{ca}. Therein, as the system's parameters vary along a closed path, the internal atomic states flip but only one way around. The parameters are chosen such that all the trapped atoms encircle a common EP in a similar fashion, leading to a collective chiral state flip.

In another more commonly discussed scenario, the jump operator is replaced by $L^\prime_1=\sqrt{\Gamma}|1\rangle \langle 2|$, meaning the spontaneous emission from state $|2\rangle$ to $|1\rangle$. Such a decay channel can be engineered in the current setup, for instance, by coupling $|2\rangle$ to an excited state which decays back to state $|1\rangle$.
One can still recover the non-Hermitian description by following the no-jump condition. But as the final state of the spontaneous emission remains in the $\{|1\rangle, |2\rangle\}$ subspace, in practice, the no-jump condition becomes more difficult to satisfy for an ensemble of atoms. Specifically, the timescale within which the non-Hermitian Hamiltonian reins decreases exponentially with the total atom number, as any single jump would leave the corresponding atom in a mixed state and beyond the non-Hermitian description.
The accumulation of these events would effectively heat up the system, and eventually make the non-Hermitian description invalid even on an approximate level. This is why such a scheme is
typically adopted for single- or few-qubit systems, but difficult to implement in atom gases.

\section{Encircling the LEP}

However, the reality is more complicated than the simple picture above. While the non-Hermiticity is introduced through the laser-assisted atom loss, the quantum jump operator $L_1$ is not the only spontaneous emission process that can happen. When the atoms are in the excited $^3P_1$ state, there is a finite probability for it to decay back to the original state ($|2\rangle$ in our convention).
After adiabatic elimination of the excited state, the process gives rise to an additional jump operator $L_2=\sqrt{\gamma}|2\rangle\langle 2|$,
which corresponds to the dephasing between states $|1\rangle$ and $|2\rangle$, due to the spontaneous emission from the excited state back to $|2\rangle$.

\begin{figure*}[tbp]
	\centering
	\includegraphics[scale=0.9]{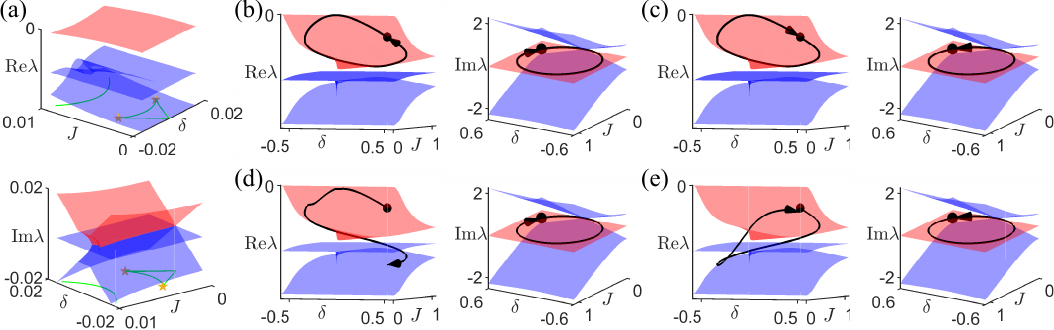}
	\caption{(a) Liouvillian eigenspectra and exceptional structure on the $\delta-J$ plane of (\ref{eq:Lcoldatom}). The green lines are the second-order Liouvillian exceptional lines and the yellow stars indicate the third-order LEP.(b)(c) Trajectories against the Liouvillian eigenspectra of (\ref{eq:Lcoldatom}) in the adiabatic limit for (b) colckwise and (c) counterclockwise rotations. (d)(e) Trajectories against the Liouvillian eigenspectra for an intermediate encircling time. The encircling path is: $\delta(t)=0.5\sin{(\pm 2\pi t/T+2\pi/3)}$, and $J(t)=0.5+0.5\cos{(\pm 2\pi t/T+2\pi/3)}$. The eigenspectra are colored red (blue), indicating the quasi-steady (excited) nature of the corresponding Liouvillian eigenstates.
We take $T=150$ for (a)(b) and $T=10000$ for (c)(d). Other parameters are $\Gamma=1/20$, $\gamma=1/100$.} \label{review4}
\end{figure*}

The dephasing process further restricts the applicability of the non-Hermitian Hamiltonian. As discussed in the previous section, under $L_1$, it is sufficient to implement a no-jump condition by detecting (post selecting) atoms that remain in the trap (and hence in the $\{|1\rangle, |2\rangle\}$ subspace).
With the addition of $L_2$, atoms that remain in the trap do not necessarily satisfy the no-jump condition.
In fact, the condition requires the complete absence of spontaneous emission from $|3\rangle$ to $|2\rangle$, which limits the validity of the non-Hermitian Hamiltonian to transient dynamics. And, similar to the case with the jump operator $L^\prime_1$, the timescale of this transient dynamics decreases exponentially with increasing atom number.

Nevertheless, the exceptional points do have consequences in the open-system dynamics. This can be revealed by analysing the Liouvillian eigenspectrum and studying the LEP encircling.
Specifically, we write the Liouvillian superoperator in a non-Hermitian matrix form
\begin{align}
	\mathcal{L}=
	\begin{bmatrix}
		\begin{array}{cccc}
			0 & iJ  & -iJ & 0 \\
			iJ &  -i\delta-\frac{\Gamma+\gamma}{2} & 0 & -iJ \\
			-iJ & 0 & i\delta-\frac{\Gamma+\gamma}{2} & iJ\\
			0 & -iJ & iJ & -\Gamma\\
		\end{array}
	\end{bmatrix},\label{eq:Lcoldatom}
\end{align}
where we enforce the no-jump condition for operator $L_1$. Physically, this corresponds to detecting atoms that remain in the trap, in the presence of dephasing dictated by $L_2$.
Similar to the example in Sec.~II, the Liouvillian eigenspectrum $\lambda$ is calculated through $\mathcal{L}\rho_\lambda=\lambda\rho_\lambda$.
But a steady-state solution with $\lambda=0$ is no longer present, due to the post selection process which does not reserve the trace of the density matrix.
We show the Liouvillian eigenspectrum in Fig.~\ref{review4}, where we identify the high-lying sheet as the quasi-steady-state solution. The quasi-steady state (with their $\text{Re}\lambda<0$) would become steady state (with $\lambda=0$) when the strength of $L_1$ is continuously turned off.
More importantly, an exceptional structure is identified in the eigenspectrum below the quasi-steady state, see for instance Fig.~\ref{review4}(a).
More concretely, with the addition of $L_2$, the HEP of the original non-Hermitian Hamiltonian develops into an exceptional structure in the Liouvillian eigenspectrum, featuring exceptional lines ending at higher-order LEPs, with a structure similar to that in Fig.~\ref{review1}.

The encircling dynamics of similar LEPs was first experimentally studied in superconducting qubits~\cite{sq1,sq2}, where chiral behaviors were reported to persist for adiabatic encircling. However, a relevant question is, since there are no exceptional structures in the steady-state subspace of these systems, the long-time dynamics should still be dominated by the steady state, and not by the exceptional structure.

The question is resolved in Ref.~\cite{sunyi}, where chiral state transfer in the presence of LEP is studied in the context of the cold-atom experiment.
First, since adiabaticity corresponds to the requirement that the rate of parameter change be much smaller than the instantaneous Liouvillian gap, the system would remain in the (quasi)-steady state of the Liouvillian. It follows that, since the LEP structure lies in the excited eigenstates of the Liouvillian, it has no bearing on the adiabatic dynamics.
Indeed, as demonstrated in Ref.~\cite{sunyi}, in the long-time limit, the system always returns to the initial state, regardless of the encircling direction. This is illustrated in Fig.~\ref{review4}(a)(b).

By contrast, a chiral state transfer is observed in the intermediate regime, in the density-matrix evolution driven by the Liouvillian superoperator. This is illustrated in Fig.~\ref{review4}(c)(d).
Qualitatively, such a chiral state transfer can be understood as the manifestation of the LEP encirlcing, which is exactly relevant at intermediate times.
We note that, the observed LEP encircling in superconducing qubits also occur at intermediate time scales~\cite{sq1,sq2,sunyi}. Similarly, a recent demonstration of the LEP-facilitated  quantum heat engine is also based on the dynamic impact of LEP at intermediate timescales~\cite{qhe}.

\section{Steady-state LEP}
Exceptional structures can also emerge in the steady-state subspace of a quantum open system. In this case, the EP-related chiral state transfer occurs in the long-time limit, even as the system adiabatically follows the steady state.
The bifurcation of the steady-state solutions at the LEP corresponds to multi-stability, which is commonly observed in non-linear systems or quantum many-body systems under the mean-field approximation~\cite{nlep4,rydbergexp}.

\begin{figure}[htbp]
	\centering
	\includegraphics[scale=1]{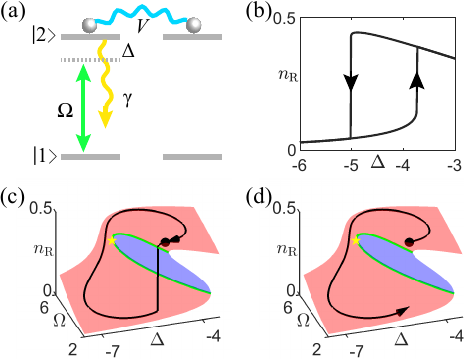}
	\caption{(a) Rydberg level scheme. (b) Steady-state solutions of $n_R=\rho_{22}$ from the optical Bloch equations under $\Omega=2$. (c)(d) Trajectories on the landscape of the steady-state Rydberg population. The green lines are the second-order Liouvillian exceptional lines and the yellow star is the third-order LEP. The red (blue) region indicates the stable (unstable) steady state.  The encircling path is: $\Omega(t)=3.85+1.477\sin{(\pm 2\pi t/T+\phi_0)}$, $\Delta(t)=-5.6+1.477\cos{(\pm 2\pi t/T+\phi_0)}$, and $\phi_0=-\arctan{(9/4)}$. Here $T=50000$, and other parameters are: $\gamma=1$, and $(N-1)V=-11$.}
\label{review5}
\end{figure}

For instance, consider the following optical Bloch equations
	\begin{align}
\dot{\rho}_{22}&=-\Omega \operatorname{Im} \rho_{21}-\gamma \rho_{22}, \label{eq:op1}\\
\dot{\rho}_{2 1}&=i\left[\Delta-(N-1)V \rho_{22}\right] \rho_{21}-\frac{\gamma}{2} \rho_{21}+i \Omega\left(\rho_{22}-\frac{1}{2}\right),\label{eq:op2}
	\end{align}
which describe the dynamics of a driven-dissipative Rydberg gas of $N$ atoms. As illustrated in Fig.~
\ref{review5}(a),  the states $|1\rangle$ and $|2\rangle$ correspond to the ground and Rydberg states, respectively. The density-dependent detuning $\Delta-(N-1)V \rho_{22}$ originates from the Rydberg interactions under the mean-field approximation~\cite{carr_2013_prl,trv1,trv2,trv3,trv5}, while $\Omega$ and $\Delta$ are the Rabi frequency and detuning of the Rydberg coupling laser.

The steady-state solutions of the optical Bloch equations are obtained by setting $\dot{\rho}=0$ and solving the resulting algebraic equations. With the non-linear detuning, the system admits either one or three steady-state solutions, depending on the parameters.
As shown in Fig.~\ref{review5}(b), in the regime with three solutions, there are an unstable solution and two stable ones.
The two stable states are both many-body steady states featuring high and low Rydberg excitations, respectively.
This is the well-known bistability which gives rise to the hysteresis in the light transmission of Rydberg gases in the electromagnetically induced transparency (EIT) measurements.

Crucially, new insights can be obtained from the corresponding Liouvillian eigenspectrum of the optical Bloch equations. As shown in Fig.~\ref{review5}(c)(d), the boundary of the bistable region marks the coalescence of the unstable and stable steady-state solutions. They can hence be identified as the second-order Liouvillian exceptional lines, which terminate at a third-order exceptional point where all three steady-state solutions merge. Here only the eigenstates of the Liouvillian superoperator coalesce, while their eigenvalues remain zero due to the steady-state nature. This is different from the EPs (HEP or LEP) in linear systems where both eigenstates and eigenvalues coalesce.
Further, since the exceptional structure in the steady states arises from non-linearity, they do not reside on the Riemann surfaces, as in the case of conventional EPs. Nevertheless, the overall topology of the exceptional structure is similar, leading to similar dynamics features as discussed below. In the literature, such exceptional structures consisting of EP lines and higher-order EPs are also called exceptional nexus~\cite{nlep3,nexus2}.

Unlike the previously discussed LEPs in the excited states, the exceptional structure in the steady-state subspace has significant impact on the long-time dynamics. An outstanding example is the chiral state transfer, as recently demonstrated in a thermal Rydberg gas~\cite{rydbergexp}. When the parameters are slowly modulated around the exceptional structure in a closed loop, the final state either switches or remains unchanged, depending on the modulation direction.
As illustrated in Fig.~\ref{review5}(c)(d), along one direction, the system adiabatically follows the steady state, whereas along the other direction, a jump is unavoidable as the trajectory crosses the exceptional line. Based on numerical simulations, a set of sufficient conditions for the chiral state transfer is summarized~\cite{rydbergexp}:
i) the initial stable steady state should be prepared in the bistable region;
ii) the two nearest (with respect to the initial position) intersection points between the trajectory and the exceptional lines should be on either side of the third-order LEP.
Under these conditions, the chiral state transfer can still be observed, even as the trajectory
crosses the exceptional lines multiple times.
Further, since the chiral state transfer only occurs when the trajectories are traversed sufficiently slowly, so that the exceptional landscape in the steady state becomes important. The chirality disappears for sufficiently fast parameter changes~\cite{rydbergexp}.
An interesting feature of the system is that both the exceptional structure and the chiral state transfer are subject to the tuning of many-body parameters. In thermal Rydberg atoms, these parameters include additional microwave fields that couple different Rydberg states, or the temperature which affects the density of the thermal gas.
These possibilities pave the way for interesting schemes of quantum control.

Note that similar exceptional structures have been reported in non-linear non-Hermitian systems~\cite{nlep1,nlep2,nlep3}. The exceptional structure therein derives from the classical non-linearity, with the  simultaneous coalescence of eigenstates and eigenvalues. It is therefore fundamentally different from the Liouvillian exceptional structure in the steady-state manifold. Nevertheless, chiral state transfer can also be observed based on the non-linear HEPs~\cite{nlep4}.

\section{Summary}
Exceptional structures and the corresponding exceptional dynamics occur in a wide range of settings. While they have attracted significant interest in classical non-Hermitian models, recent studies have revealed their relevance and impact in quantum open systems, where exceptional structures emerge in the Liouvillian eigenspectrum and affect the steady-state approaching dynamics.
Our brief review focuses on the chiral state transfer, a particular dynamic consequence, near the Liouvillian EPs.
Based on these studies, it would be interesting to explore other exceptional features such as the EP-enhanced sensitivity and criticality near the Liouvillian EPs in quantum open settings.
This is particularly intriguing in physical platforms such the driven-dissipative Rydberg gases, where the development of novel quantum control and sensing schemes would also benefit the on-going explorations therein for quantum information and computation.
It is also interesting to study the generation and dynamic consequences of higher-order LEPs, wherein the complicated exceptional landscape can lead to richer encircling possibilities but have rarely been explored in quantum open systems.
Finally, given the recent interest in the non-Abelian braid topology in non-Hermitian multi-band systems~\cite{nonabelian1,nonabelian2}, it is tempting to investigate and exploit similar features in quantum open systems, based on our understanding of the LEPs.

{\bf Acknowledgement}

See Funding section below.

{\bf Funding}

This research is supported by the Natural Science Foundation of China (Grant No. 12374479).

{\bf Author contribution}

K. S. performed theoretical derivations and numerical calculations.
W. Y. supervised the project. Both authors contributed to the writing of the manuscript.

{\bf Availability of Data and Materials}

Codes and numerical results of this review are available from the corresponding author upon reasonable request.

{\bf Competing interest declaration}

The authors declare no competing interests.

\end{document}